\documentclass[useAMS,usenatbib]{mn2e}

\usepackage{graphics}
\usepackage{epsf}
\usepackage{subfigure}

\title[PC \& AC Relations VI]{Period-Colour and Amplitude-Colour Relations in Classical Cepheid Variables - VI. New Challenges for Pulsation Models}

\author[Kanbur et al.]{Shashi M. Kanbur$^{1}$\thanks{E-mail: shashi.kanbur@oswego.edu}, M. Marconi$^{2}$, C. Ngeow$^{3}$, I. Musella$^{2}$, \and M. Turner$^{4}$,
and A. James$^{5}$, and S. Magin$^{1}$, and J. Halsey$^{1}$
\\
$^{1}$Department of Physics, State University of New York at Oswego, Oswego, NY 13126, USA
\\
$^{2}$Osservatorio Astronomico di Capodimonte, Via Moiariello 16, 80131 Napoli, Italy 
\\
$^{3}$Graduate Institute of Astronomy, National Central University, Jhongli City, 32001, Taiwan (R.O.C.)
\\
$^{4}$Department of Physics, Rice University, USA
\\
$^{5}$Department of Physics, State University of New York at Geneseo, Geneseo, NY, USA
}

\begin{document}

\date{Accepted 2010 month date. Received 2010 month date; in original form 2010 month date}

\pagerange{\pageref{firstpage}--\pageref{lastpage}} \pubyear{2010}

\maketitle

\label{firstpage}

\begin{abstract}

We present multiphase Period-Color/Amplitude-Color/Period-Luminosity relations using OGLE III
and Galactic Cepheid data and compare with state of the art theoretical pulsation models.
Using this new way to compare models and observations, we find convincing evidence that both
Period-Color and Period-Luminosity Relations as a function of phase are dynamic and highly nonlinear at certain pulsation
phases. We extend this to a multiphase Wesenheit function and find the same result. Hence our results cannot be due
to reddening errors. We present statistical tests and the urls of movies depiciting the Period-Color/Period-Luminosity and Wesenheit
relations as a function of phase for the LMC OGLE III Cepheid data: these tests and movies clearly demonstrate nonlinearity as a function of phase
and offer a new window towards a deeper understanding of stellar pulsation.
When comparing with models, we find that the models also predict this nonlinearity in both Period-Color
and Period-Luminosity planes. The models with $(Z=0.004,Y=0.25)$ fare better in mimicking the LMC Cepheid relations, particularly
at longer periods, though the models predict systematically higher amplitudes than the observations. 

\end{abstract}


\section{Introduction}

Classical Cepheids are the most important primary distance indicators
within the Local Group thanks to their characteristic
Period-Luminosity (PL) and Period-Luminosity-Color (PLC)
relations. They are also currently used to calibrate secondary
distance indicators and in turn to evaluate the Hubble constant
\citep[$H_0$, see e.g.][]{fre01,sa01}. Any systematic effect on the
Cepheid PL relations is expected to propagate on secondary distance
indicators and, in turn, on the final evaluation of $H_0$. During the last decade
there has been a very lively debate on the possible non-universality of
Cepheid PL relations traditionally assumed to have the slope of the
one derived in the Large Magellanic Cloud \citep[see e.g.][and
references therein]{fre01,b08,m09}. Unfortunately, in spite of many 
observational and theoretical efforts, there is not a conclusive answer 
to this question: nonlinear pulsation models predict a significant 
dependence both on metallicity and the helium content with final effects on 
the distance scale that can be higher than 10 \% \citep[][and references therein]{m09,b08} and empirical tests that either
find no metallicity effect or predict a metallicity dependence with an opposite sign with respect to the theoretically
predicted one. However, spectroscopically based analyses by \citet{r05,r08} support the trend suggested by the
pulsation model results.

Beyond the debated dependence on chemical composition there are both
observational \citep[e.g.][]{kn04,n05,n09} and theoretical
\citep[e.g.][]{cmm00,mmf05,knb04} indications that the Cepheid PL
relation is not always linear. Ngeow \& Kanbur (2006a) have also shown that the
effect on the inferred $H_0$ of neglecting this possible nonlinearity
can be of $1-2 \%$, that combined with the predicted dependence on
chemical composition can be important in the light of current efforts
to push down errors on the $H_0$ estimate to a few percent.

The Cepheid
PL relation used for distance determination is evaluated at mean
light: the mean over all pulsation phases. On the other hand recent
studies by \citet{kn04,n05,n06} have suggested that an innovative and
promising approach to Cepheid pulsation is investigating the
Period-Color (PC) and the PL relations as a function of phase. Indeed
modern data on Cepheids are characterized by excellent phase
coverage, particularly in the Magellanic Clouds. This offers a
unique opportunity for deriving empirical multiphase PC and PL
relations. This approach has already been adopted by \citet{n06} on the basis
of data from the OGLE II project for Magellanic Cepheids and from various literature sources for Galactic pulsators. These authors have demonstrated
that the study of
multi-phase relations can provide new insights into Cepheid pulsation. They find strong evidence that these
observed PC/PL relations vary significantly with phase, are highly dynamic and strongly nonlinear at various pulsation phases usually
located near minimum light. These relations offer a new way to constrain models and provide deeper insights
into pulsation physics by comparing these observed multiphase
PC/PL relations with those from theoretical models.
In this paper we update our multiphase PC/PL relations using OGLE III LMC data and present a detailed
comparison between theoretical and empirical multiphase PC and PL
relations for the Milky Way and the Large Magellanic Cloud (LMC).

\citet{skm93}, \citet{kn04} and \citet{n06} also showed the importance of amplitude-color (AC) relations. A flat
PC relation at maximum/minimum light implies a relation between amplitude and color at minimum/maximum light. \citet{kn04} found evidence
of a change of slope in the PC relation at maximum light at a period of about 10 days and a corresponding change in the AC relation. Hence
this paper also presents AC relations as a function of phase.

\section{The Observed Multiphase PC/PL Relations}

The data used in this study comes primarily from OGLE III and were
corrected for reddening using the methods described in \citet{n09}.
The data were fitted with a Fourier series like
$$V = A_0 + \sum_{k=1}^{k=N}A_k\sin(k{\omega}t + {\phi}_k),$$
where $A_0$ is the mean magnitude, ${\omega} = {{2\pi}\over{P}},$ and $P$ is the period in days, $t$
is the time of observation and $N$ is the order of the fit. Because the phase coverage and the number of
stars is large, the order of the Fourier fit was 8. The resulting Fourier fit was then used to
infer properties at maximum/minimum/mean light. We performed a number of tests which checked that the observed
multiphase PC/AC/PL relations were independent of the order of the Fourier fit.
The Fourier fits were rephased so that maximum light occurs at phase 0 in the V band, as described in
\citet{n06}.

In what follows, short and long period Cepheids are demarcated by a period of 10 days.
Our results are presented in table 1 and figures 1-8 using OGLE III data for the LMC results together with results
for Galactic Cepheids using data from \citet{n06}.
Following an anonymous referee's suggestions we present only figures 1 and 2 in the published version. Figures 3-8 are 
in the on-line version of the paper. However in the text that follows we do refer to all these figures.

In all these figures, short/long period observed data are represented by
black and blue crosses respectively.
Theoretical results are usually
represented by colored symbols related to the mass of the models in question. These are noted on the figures.
Some of our results pertain solely to the observed behavior, some
solely to the theoretical relations and some to the comparison of both. However, to save space,
we have superimposed the theoretical relations to the observed ones.

The results presented in figures 1-8 depict PC/AC/PL relations at maximum and minimum light.
These figures broadly support the work of \citet{n06} and clearly demonstrate the dynamic nature of the PC/PL
relations as a function of phase. These figures, and the movies (see also the url addresses given below),
provide clear evidence of a nonlinearity at phases close to minimum light - statistical
tests are not needed. This nonlinearity is seen in PC/PL/AC relations: changes in one relation are reflected in the
other.
In particular, we see that at minimum light, the nonlinearity is marked, and moreover, it is sharp: see also
\cite{n06} and figures 5 and 6 here. That is the data
strongly imply a significant change in the slope of both PC/AC/PL relations at a period of about 10 days.

Table 1 presents the results of $F$ tests as outlined in \citet{n06} and
references therein for the PC and PL relations at maximum and minimum 
light for the LMC using OGLE III Cepheids. In this table, we present the slope and zero points for 
the long and short period regression lines, the dispersion, the number of data points used and the
value of the $F$ statistic. For the number of data points used here, an $F$ value greater than about 3 indicates that the
data are more consistent with the alternative hypothesis at the $95\%$ confidence interval.
We note that certain phases such as minimum
light are strongly non-linear according to the $F$ test whilst other phases (close to maximum
light) are marginally linear. The combination of these phases again results in a dilution of the nonlinearity
at mean light and suggests why this effect has not been observed before. Note that \citet{n09} only presented the results of such statistical tests
at mean light.

\begin{table*}
\centering
\caption{$F$ test results for PC/PL/PW relations at various phases}
\label{NOLABEL}
\begin{tabular}{ccccc}\hline
Period&Slope&ZP&${\sigma}$&N\\
\hline
$LMC PC(max)$\\
\hline
All&$0.241\pm 0.018$&$0.296\pm 0.012$&0.144&1625\\
$\log(P)<1.0$&$0.319\pm0.026$&$0.251\pm0.016$&0.142&1517\\
$\log(P)>1.0$&$-0.246\pm0.134$&$0.830\pm0.156$&0.159&108\\
$F$-statistic is 14.87 (non-linear)\\
\hline
$LMC PC(min)$\\
\hline
All&$0.281\pm 0.015$&$0.573\pm0.01$&0.117&1602\\
$\log(P)<1.0$&$0.182\pm-0.021$&$0.629\pm0.013$&0.114&1494\\
$\log(P)>1.0$&$0.510\pm0.113$&$0.352\pm0.131$&0.134&108\\
$F$-statistic is 22.255 (non-linear)\\
\hline
$LMC PL_V(max)$\\
\hline
All&$-2.672\pm0.044$&$16.691\pm0.030$&0.351&1624\\
$\log(P)<1.0$&$-2.574\pm0.064$&$16.635\pm0.039$&0.351&1516\\
$\log(P)>1.0$&$-3.032\pm0.291$&$17.066\pm0.337$&0.344&108\\
$F$-statistic is 2.71 (linear)\\
\hline
$LMC PL_V(min)$\\
\hline
All&$-2.639\pm0.038$&$17.395\pm0.025$&0.298&1621\\
$\log(P)<1.0$&$-2.883\pm0.053$&$17.535\pm0.033$&0.293&1513\\
$\log(P)>1.0$&$-1.658\pm0.260$&$16.365\pm0.301$&0.307&108\\
$F$-statistic is 24.87 (non-linear)\\
\hline
$LMC PW(phase=0.0)$\\
\hline
All&$-3.279\pm0.020$&$15.921\pm0.013$&0.160&1629\\
$\log(P)<1.0$&$-3.368\pm0.027$&$15.973\pm0.017$&0.151&1522\\
$\log(P)>1.0$&$-2.437\pm0.195$&$14.976\pm0.226$&0.229&107\\
$F$-statistic is 26.16 (non-linear)\\
\hline
$LMC PW(phase=0.25)$\\
\hline
All&$-3.321\pm0.027$&$15.819\pm0.013$&0.152&1629\\
$\log(P)<1.0$&$-3.252\pm0.027$&$15.780\pm0.016$&0.146&1522\\
$\log(P)>1.0$&$-3.231\pm0.178$&$15.677\pm0.207$&0.209&107\\
$F$-statistic is 6.97 (non-linear)\\
\hline
$LMC PW(phase=0.5)$\\
\hline
All&$-3.275\pm0.0233$&$15.793\pm0.014$&0.172&1629\\
$\log(P)<1.0$&$-3.199\pm0.030$&$15.750\pm0.019$&0.166&1522\\
$\log(P)>1.0$&$-3.465\pm0.194$&$15.977\pm0.225$&0.228&107\\
$F$-statistic is 6.28 (non-linear)\\
\hline
$LMC PW(phase=0.75)$\\
\hline
All&$--3.329\pm0.028$&$15.921\pm0.019$&0.223&1629\\
$\log(P)<1.0$&$-3.391\pm0.040$&$15.956\pm0.025$&0.222&1522\\
$\log(P)>1.0$&$-3.121\pm0.199$&$15.708\pm0.230$&0.233&107\\
$F$=statistic is 2.61 (linear)\\
\hline

\end{tabular}
\end{table*}

Our results support and extend the conclusions of \citet{n06}
\begin{itemize}
\item There is a marked nonlinearity at a period of $\log P \approx 1$ in the PC/PL/AC plots.
The PL relation at maximum light is linear and is strongly nonlinear at minimum light.
\item The PC plots show a great deal of structure with, possibly, changes of slope at other periods as well as at $\log P \approx 1.$
\item There is a variation in dispersion of both PC/PL relations as a function of phase. The PC and PL relation at maximum light has
a greater dispersion than at minimum light. Moreover the dispersion for a given phase also decreases at periods close to 10 days - at
least for the LMC.
\item Both Galactic and LMC PC relations at maximum light are flat but the OGLE III results again suggest that the Galactic PC relation is
flat for $\log P \ge 0.8$ whilst the LMC PC relations is flat for $\log P \ge 1$.
\item We see that at minimum light, higher amplitude Cepheids are driven to redder colors when the PC relation is
flat or flatter for those periods - as originally proposed by \citet{skm93}. For example, the AC(min) relation
for Galactic Cepheids is one relation across the entire period range. However, the AC(max) relation is clearly two
separate relations, demarcated by a sharp change at a period of 10
days with the longer period relation having a non-zero slope \citep{n06}. This corresponds to a change in the slope of the LMC PC(max and min) relation at a period of 10 days. This is important because a number of authors
\citep{kw06} have noted correlations between mean color and amplitude. Since mean light relations are simply the average of
the same relations as a function of phase, then a correlation between mean color and amplitude has to be due to the correlation between
amplitude and pulsation phases around minimum light. This, in turn, is caused by the interaction of the hydrogen ionization front (HIF) and photosphere
\citep{skm93,knb04,kn06}. This again demonstrates why such
a multiphase analysis is useful and demanded by the high quality data now available.
\item Movies of the multiphase PC/PL relations, viewable at

\centerline{http://www.oswego.edu/$\sim$kanbur/IRES2009/Cphase.mov,}

\centerline{http://www.oswego.edu/$\sim$kanbur/IRES2009/Vphase.mov,}

provide further strong evidence of
the nonlinear nature of the PC/PL relation at phases around minimum light. Further studies of such movies is warranted since
they seem to indicate the presence of a number of shocks and different behavior for Cepheids in different period ranges.
These movies also suggest that a group of Cepheids mostly with periods
around 10 days (perhaps the bump Cepheids)
are the first to brighten subsequent to initial
dimming after maximum light. While this may be due to the Hertzpsrung progression, this offers an opportunity to study the
Hertsprung progression from a different perspective.
\end{itemize}

\section{The Theoretical Scenario}

We selected various sets of nonlinear convective pulsation models
computed with the code originally developed by \citet{s82,bs94} and
adapted to Cepheid pulsators by \citet{bms99}. 
The selected models cover a range of chemical compositions, namely Z=0.02, Y=0.31; Z=0.02, 
Y=0.28; Z=0.01, Y=0.26; Z=0.004, Y=0.25; Z=0.008, Y=0.25, 
considered representatives of Galactic and Magellanic Cepheids \citep[see][for details]{mmf05,b08,m09}. 
For each selected chemical composition and stellar mass we considered canonical models, that are models following the mass-luminosity (ML) relation derived by \citet{b00} without taking into account either overshooting or mass-loss. 
The model bolometric light curves presented in \citet{bcm00,mmf05} are converted into magnitude and color variations using
static stellar atmospheres \citep{cgk97a,cgk97b} and used to derive theoretical multiphase PC/PL relations in the various bands.
Figures 1-8 display these theoretical relations. In each plot, a given symbol represents pulsation models at a given mass (and hence luminosity,
through the adopted $ML$ relation), composition and a range of effective temperatures.

In looking at just the theoretical PC relations, we see that some mass sequences are monotonic, some suffer
a gradual change of slope whilst others are distinctly non-monotonic, and further, this behaviour is sometimes different (for
a given mass sequence)
at maximum and minimum light: for example, see figure 7 - the $9M_{\odot}, Y=0.25, Z=0.004$ mass sequence.

Examples of the non-monotonic behaviour typically bracket a period of 10 days but in some cases, this behaviour also occurs
at periods greater than 10 days.

\section{The multiphase PC/PL relations: theory versus observations}

Inspection of Figures 1-8 suggests that:

\begin{itemize}
\item The models with $Z=0.004, Y=0.25$ fare the best in reproducing the LMC PC/PL/AC relations, particularly at long periods -
see for example, figures 6-7, which compare the PC/PL relations at maximum/minimum light with model results. In particular note the comparison 
for PC relations at maximum/minimum light for periods in the range between $1.2 < \log P < 1.5.$
\item The observed LMC AC relation displays a group of stars across a wide period range which are distinct from the main group eg. figure 7.
These are modeled quite well by the 9-11 solar mass models with $(Z=0.004,Y=0.25)$ composition. So it could be that the scatter in these plots is
real and these stars are high mass stars with increasingly lower amplitude: perhaps because they are close to the edge of the instability strip.
Certainly, the non-monotonic nature of the purely theoretical AC relations at maximum/minimum light would indicate the presence of such stars. 
\item It is also interesting to note that the LMC data are also quite nicely reproduced by models with $Z=0.02$ $Y=0.31$,
whereas they disagree with models at $Z=0.02$ $Y=0.28$. This occurrence confirms the
result that helium and metallicity effects tend to compensate each other.
\item $V_{max}$ and $V_{min}$ as a function of amplitude are systematically higher for models than for observations.
This might be due to the sensitivity of model amplitudes to residual uncertianties in the treatment of turbulent convection.
\item Figures 2-4, particularly the PC relation at maximum light,
indicate that models with $(Z=0.02, Y=0.28)$ fare better at reproducing the Galactic observations than models
with $(Z=0.01, Y=0.26).$
\item For Galactic Cepheids, the amplitude range covered by models is much wider than the observed one.
\item For a given mass and periods shorter than 10 days, the theoretical PC/PL relation at maximum light has
a greater positive slope than at minimum light (where the relations are almost flat). If the models can be regarded as
doing a reasonable job of mimicking the observations, this shows why, in part, the observed
PC/PL relations have a greater dispersion at maximum rather than at minimum light - see, for example figure 7.
\end{itemize}

\section{The Multiphase Wesenheit Function}

A number of different formulations of the Wesenheit function exist in the literature.
Madore and Freedman (1998) define it to be
$$W_V = V - 2.45(V-I),$$
whilst Udalski et al (1999), who adopt a slightly different extinction
law, use the following definition,
$$W_I = I - 1.55(V-I).$$

Here the quantities $V,I$ are the mean observed magnitudes in these bands. $W_V$ and $W_I$ can be shown to be
reddening independent and for this reason the Wesenheit function is the preferred way to use the Cepheid PL
relation to estimate distances. It is of interest here to consider the linearity/nonlinearity of the Wesenheit function as
defined by Udalski et al (1999).

If we phase both $V$ and $I$ observations relative to a common time origin and then, following a Fourier decomposition, rephase
such that maximum $V$ band brightness is called phase 0, then we can formulate a multiphase "Wesenheit-type" function as
$$W_{Iph} = I_{ph} - 1.55(V_{ph} - I_{ph}),$$
where $V_{ph}$ and $I_{ph}$ denote these quantities at the phase value $ph.$
The results of the $F$ test for nonlinearity applied to the Wesenheit function are presented in table 1 and figure 9. We see clearly that
certain phases of the Wesenheit function, for example at maximum light, are strongly nonlinear. Since the reddening along the
line of sight cannot vary due to pulsation phase, this result provides strong evidence against the hypothesis that previous statistical
tests indicating nonlinearity are due to reddening errors. Previous work such as \citet{k07} suggested that linear Wesenheit functions
were due to nonlinearities in the PC and PL relations canceling out: the results presented here support this view.

Another nice demonstration of the nonlinearity of the Wesenheit function at certain phases is presented at the following url.

\centerline{http://www.oswego.edu/$\sim$kanbur/IRES2009/Wphase.mov,}

where we have 100 phase points , the $y$ axis is the Wesenheit function and the $x$ axis is $\log P.$ Again we clearly see 
a dynamic $W_I$ function which changes as a function of phase, not as much as the $V$ and $I$ band PL relations but nevertheless
the nonlinearity at certain phases is clear and unambiguous and cannot be due to lack of data in certain period ranges. We get similar results
we use the Madore and Freedman (1998) formulation.

It is our contention that the $F$ test results presented in Table 1
provide very strong evidence for a change in the slope of the PL, PC {\bf and PW} relations at a period of
10 days and provides evidence of the pulsation phases which are the most nonlinear.

\section{Conclusions}

Comparing observations and theory on the multiphase PC/PL/AC plane as described here is a powerful new way to constrain
models and gain deeper insights into pulsation physics. The dynamic nature of the multiphase plots provides a new window into
the inner workings of Cepheids and deserves more attention both observationally and theoretically. This can best be seen by viewing movies of
these multiphase PC/PL relations available - see above for web links.
The Wesenheit movie depicts the variation with phase of the Wesenheit function, defined by
$W = I - 1.55*(V-I).$ A number of authors have suggested that because previous work testing for possible nonlinearities in the Wesenheit function
has yielded negative results, indicating nonlinearities in the PL/PC relations at mean light may be due to reddening errors (the
Wesenheit function is reddening independent). Our movies indicate clear nonlinearities in the 
multiphase Wesenheit function implying that the cause of the nonlinearity cannot be due to reddening. 
The effect of this nonlinearity on the extra-galactic distance scale and CMB independent estimates of $H_0$ is a matter for debate.

It may be argued that a comparison of observations and theory in the multiphase PC/PL planes are another projection of the comparison
of observed and theoretical light curves: even if the PL/PC/PW relations at mean light were linear, there is no similar 
expectation for linearity of the multiphase relations because of, for example, the Hertzsprung progression. This
broad argument may be true but it still needs to be demonstrated and the demonstration of the dynamic nonlinear
behavior of the PL and PW relations is one of the main results of this paper. The average of the multiphase relations does
surely yield information about the mean light relations.

Does such
a comparison provide additional insight over and above a comparison of light curves? We would argue that it does. For example,
flat PC relations at maximum light yield information about the interaction of the photosphere and hydrogen ionization front - something
which would be very difficult to probe by just a comparison of observed and theoretical light curves. Our multiphase
comparisons suggest that the greatest nonlinearity occurs at minimum light: this information, when investigated in greater detail can
provide a deeper understanding of Cepheid pulsation over and above a comparison of observed and theoretical light curves. This will be the
topic of future work.

In any case the results presented in this paper represent an
important challenge for theoreticians seeking a deeper understanding of Cepheid pulsation.

\section*{Acknowledgments}
MM and IM acknowledge financial support from PRIN-INAF2008. CCN thank the funding from National Science Council (of Taiwan) under the contract NSC 98-2112-M-008-013-MY3. MT and AJ acknowledge funding from the NSF under contract number OISE-0755645 and the 2006 Chertien Award of the American Astronomical Society.
JH and SM thank SUNY Oswego for support.

\begin{figure*}
  \vspace{0cm}
  \hbox{\hspace{0.5cm}
    \epsfxsize=8.0cm \epsfbox{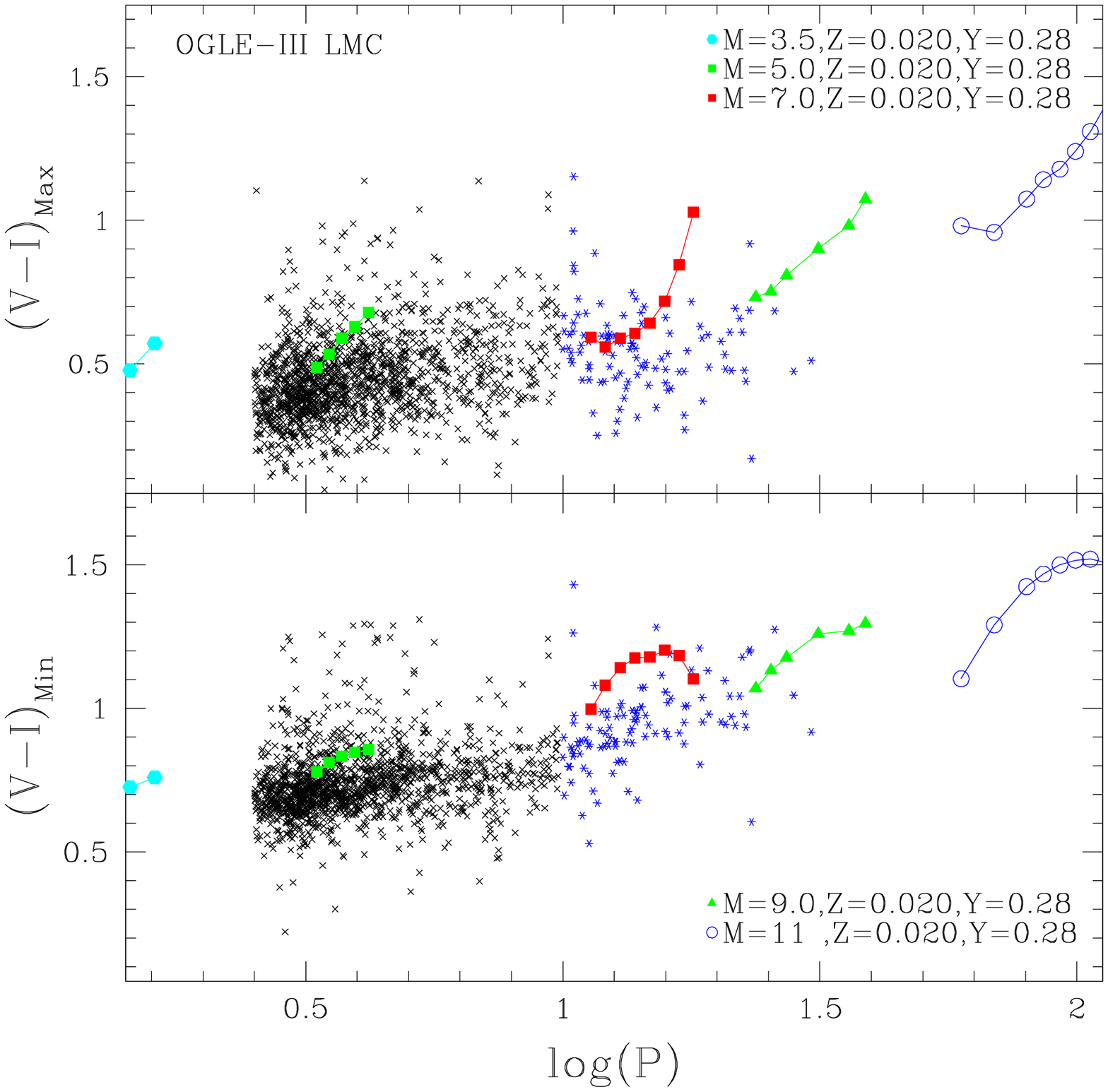}
    \epsfxsize=8.0cm \epsfbox{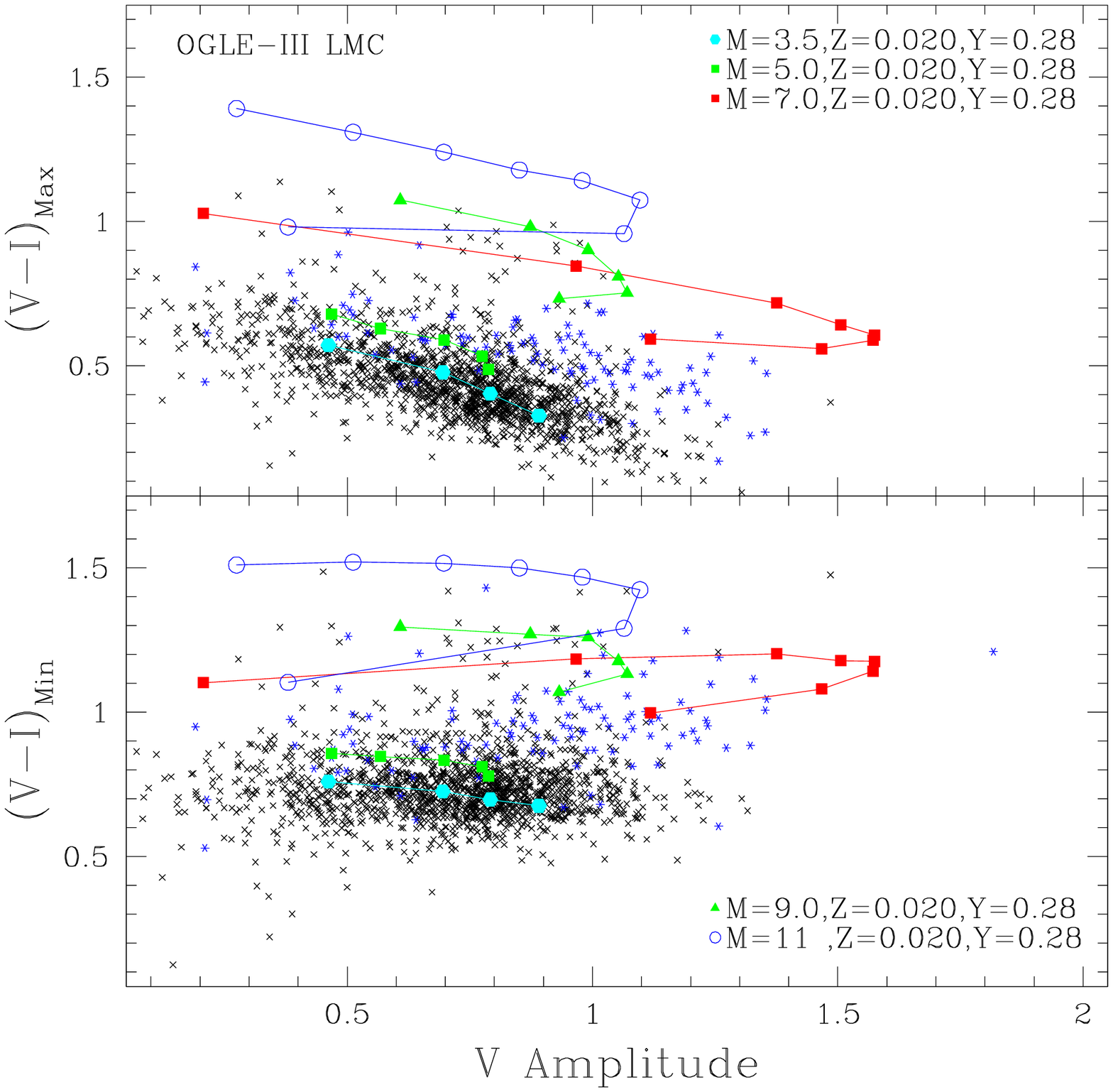}
  }
  \vspace{0cm}
  \caption{Multiphase PC/AC relations for reddening corrected OGLE III LMC Cepheid data and theoretical models with $Z=0.02, Y=0.28.$}
  \label{UNIQUE_LABEL}
\end{figure*}

\begin{figure*}
  \vspace{0cm}
  \hbox{\hspace{0.5cm}
    \epsfxsize=8.0cm \epsfbox{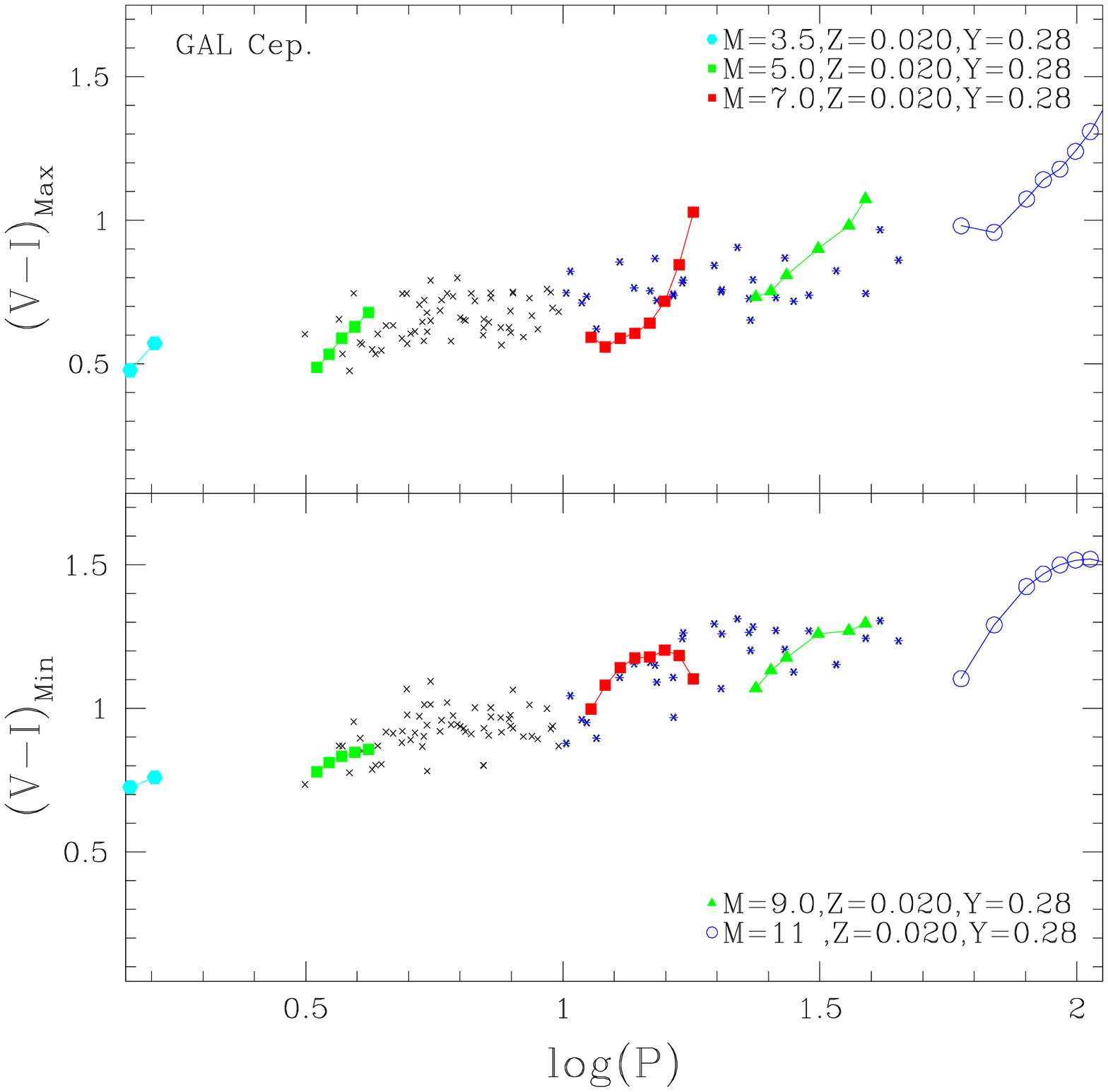}
    \epsfxsize=8.0cm \epsfbox{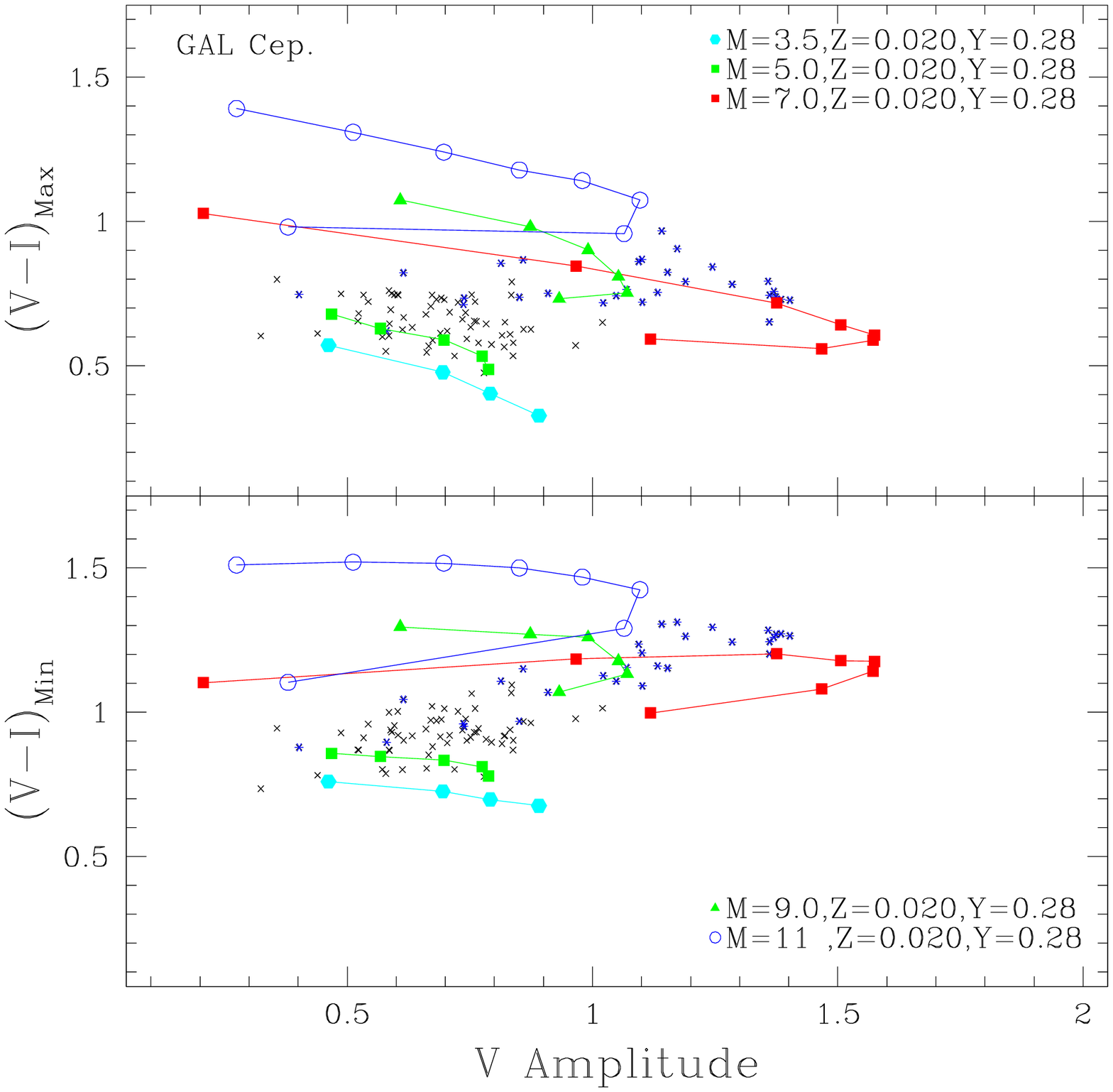}
  }
  \vspace{0cm}
  \caption{Multiphase PC/AC relations for reddening corrected Galactic Cepheid data and theoretical models with $Z=0.02, Y=0.28.$}
  \label{UNIQUE_LABEL}
\end{figure*}

\begin{figure*}
  \vspace{0cm}
  \hbox{\hspace{0.5cm}
    \epsfxsize=8.0cm \epsfbox{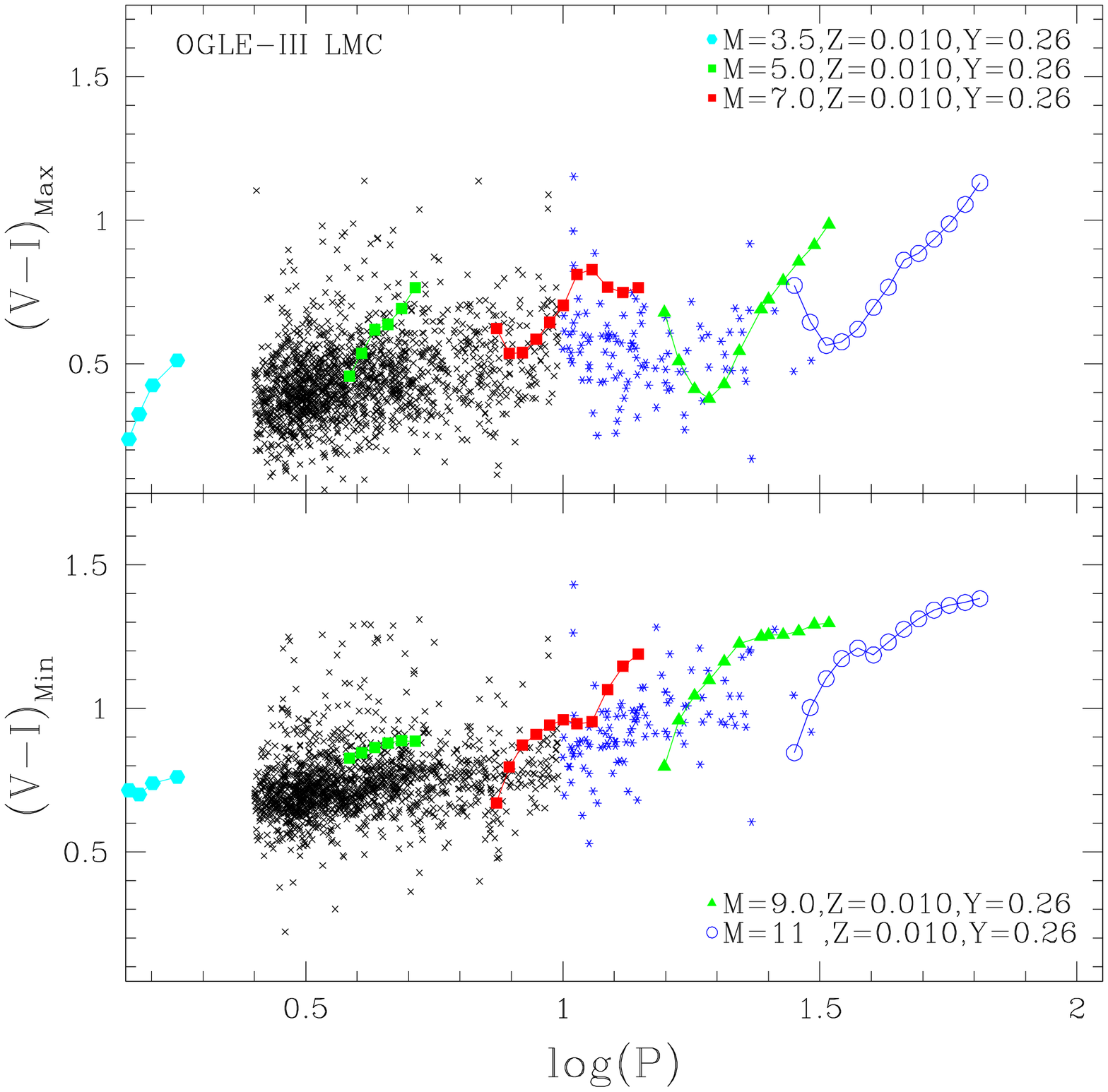}
    \epsfxsize=8.0cm \epsfbox{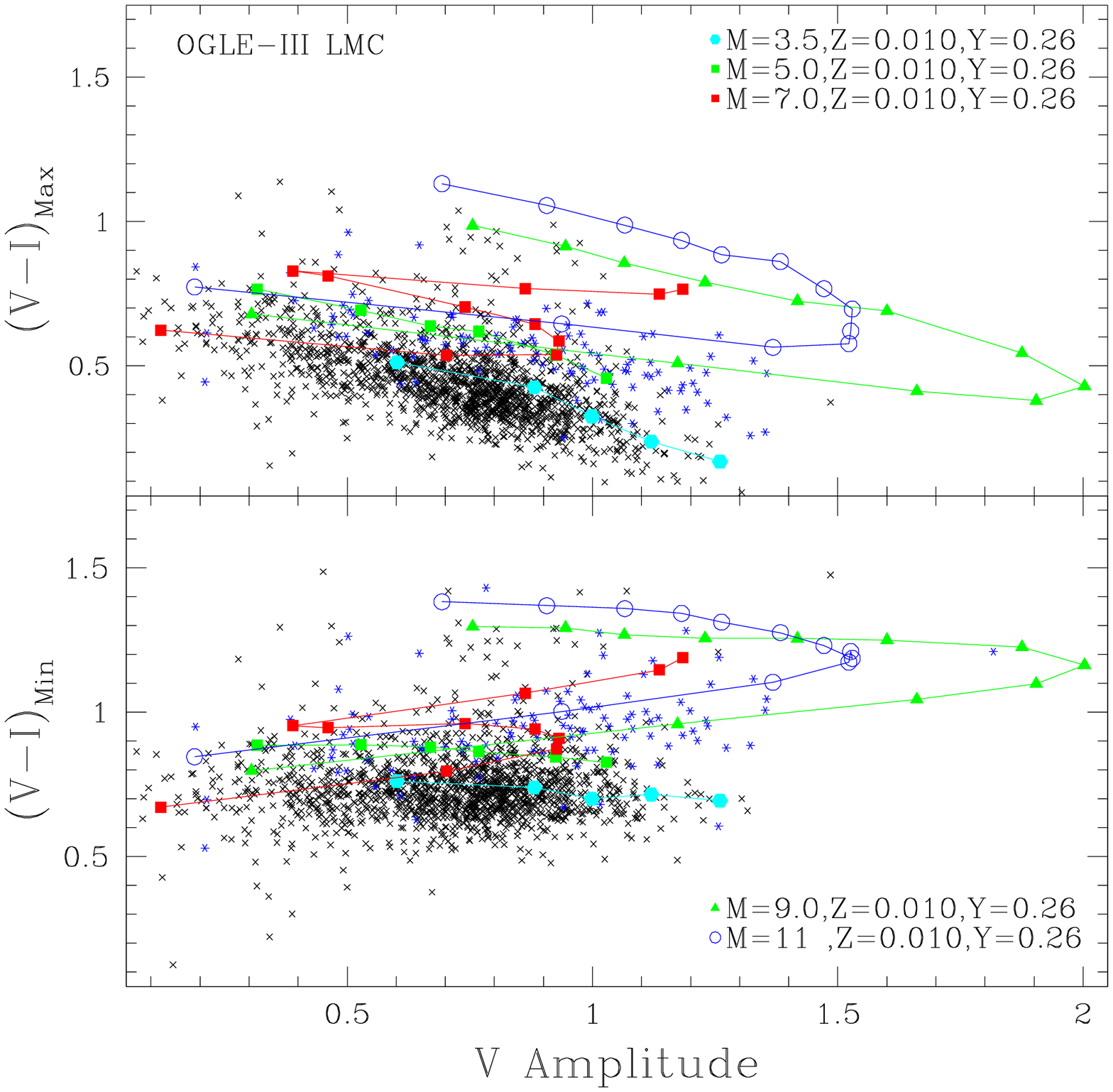}
  }
  \vspace{0cm}
  \caption{Multiphase PC/AC relations for reddening corrected OGLE III LMC Cepheid data and theoretical models with $Z=0.01, Y=0.26.$}
  \label{UNIQUE_LABEL}
\end{figure*}

\begin{figure*}
  \vspace{0cm}
  \hbox{\hspace{0.5cm}
    \epsfxsize=8.0cm \epsfbox{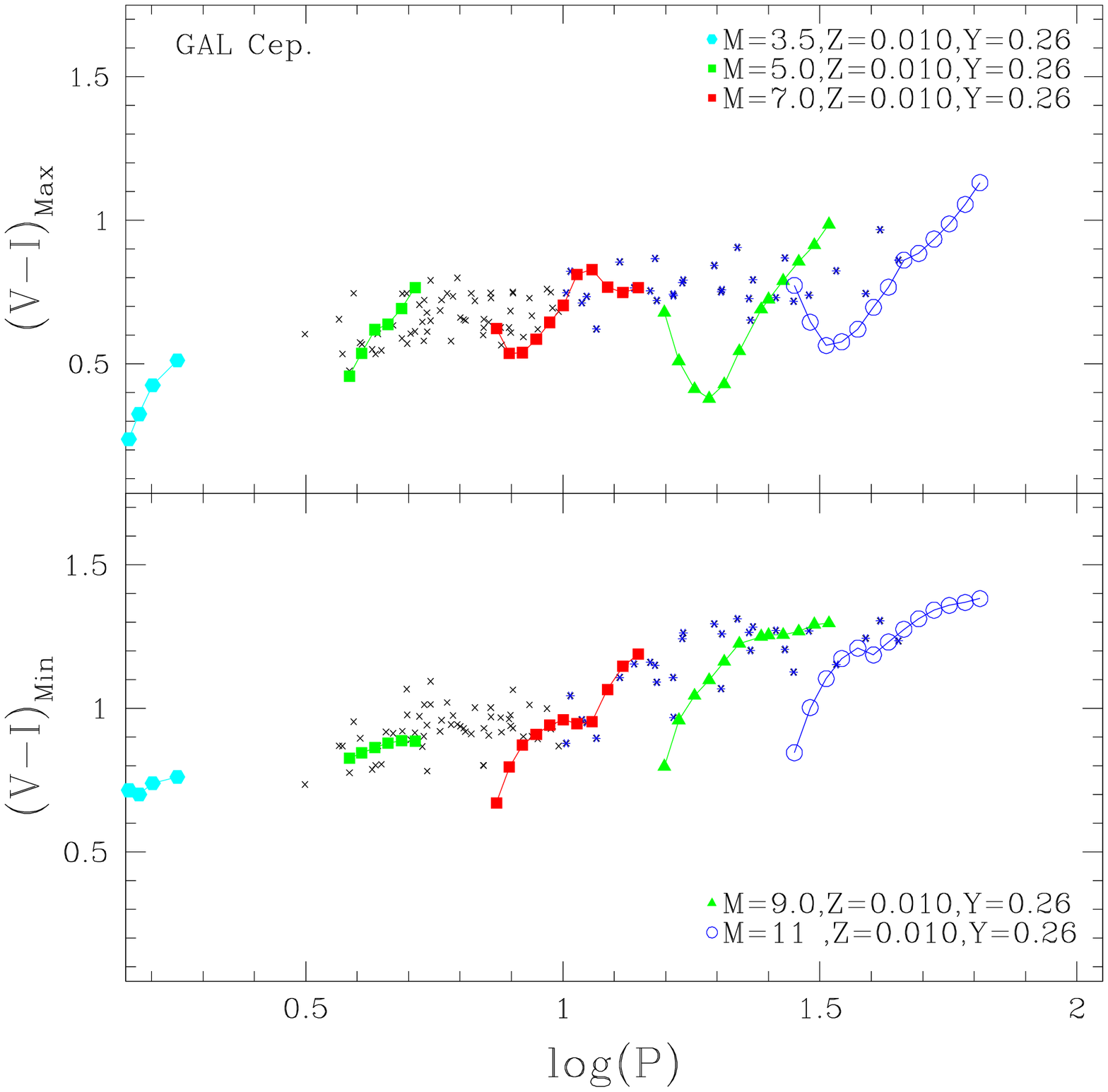}
    \epsfxsize=8.0cm \epsfbox{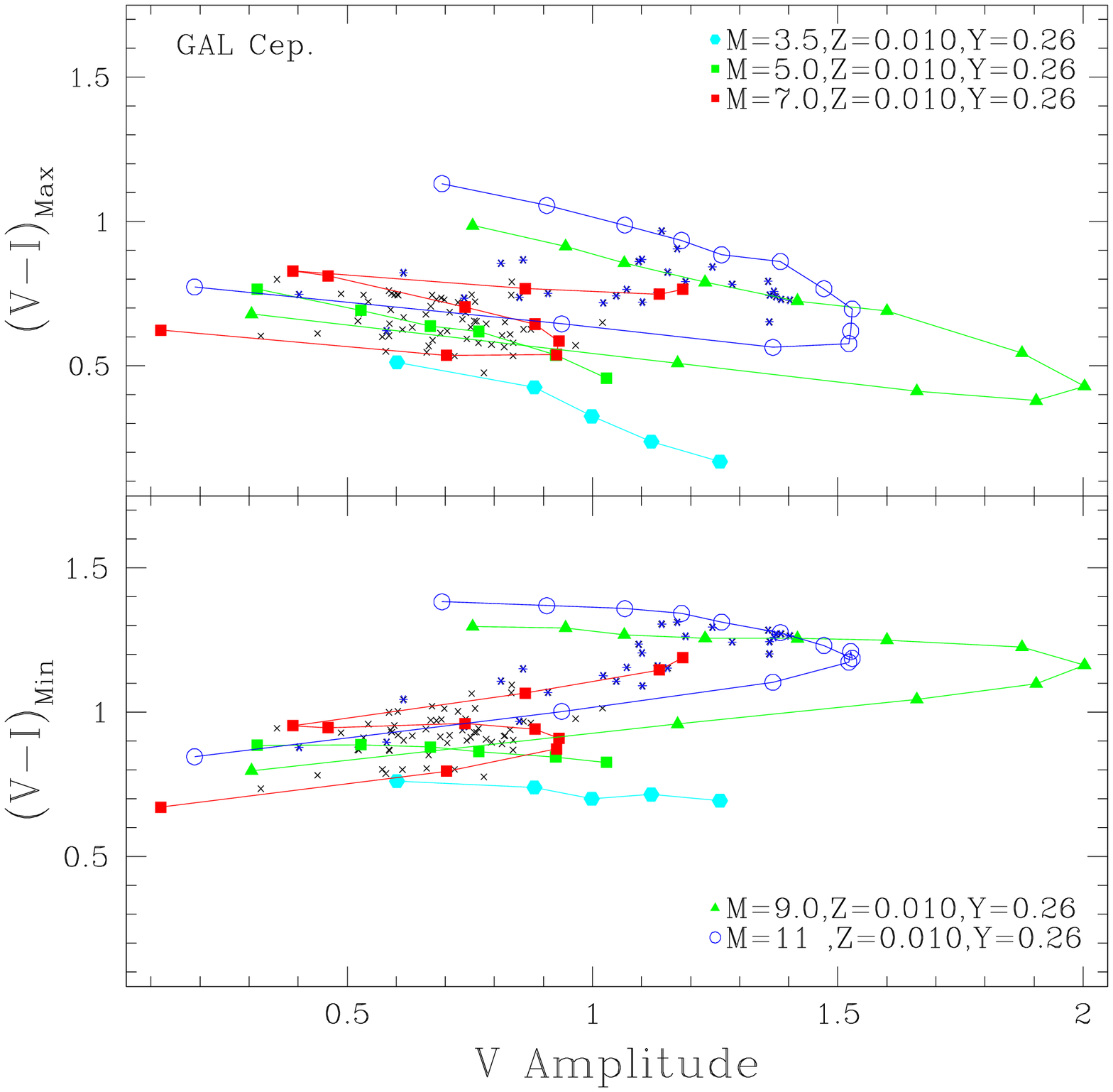}
  }
  \vspace{0cm}
  \caption{Multiphase PC/AC relations for reddening corrected Galactic Cepheid data and theoretical models with $Z=0.01, Y=0.26.$}
  \label{UNIQUE_LABEL}
\end{figure*}

\begin{figure*}
  \vspace{0cm}
  \hbox{\hspace{0.5cm}
    \epsfxsize=8.0cm \epsfbox{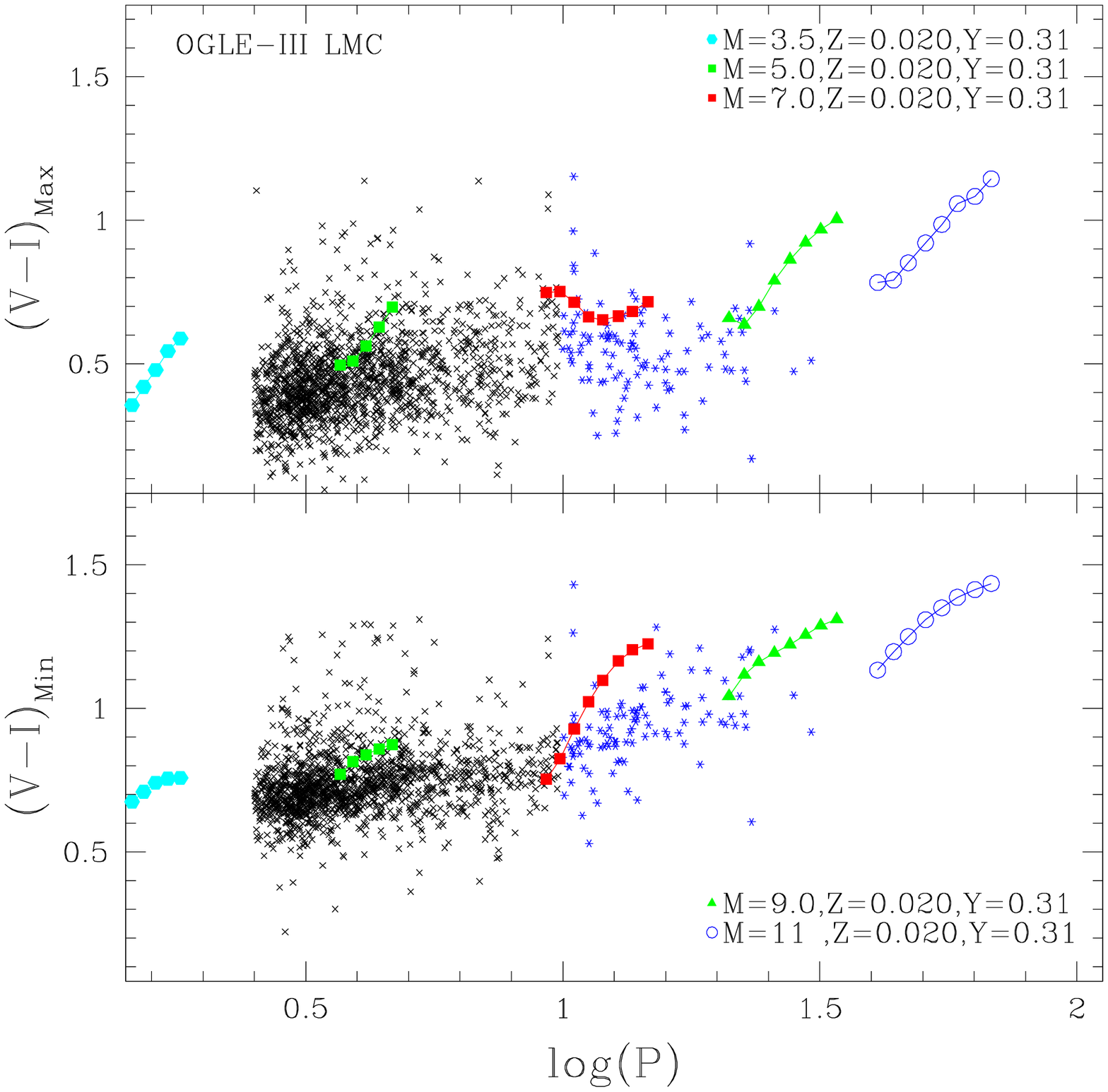}
    \epsfxsize=8.0cm \epsfbox{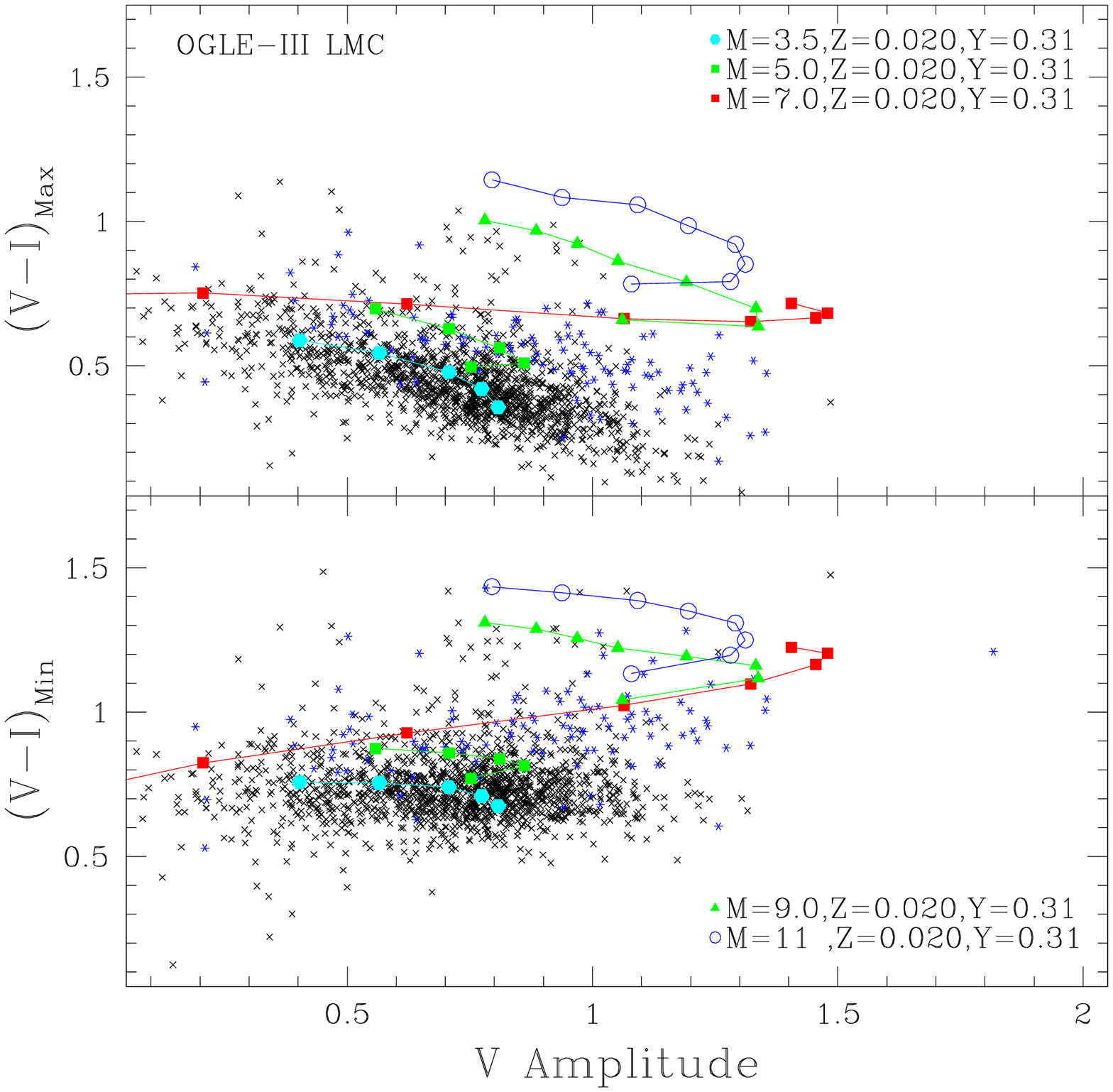}
  }
  \vspace{0cm}
  \caption{Multiphase PC/AC relations for reddening corrected OGLE III LMC Cepheid data against theoretical models with $Z=0.02, Y=0.31.$}
  \label{UNIQUE_LABEL}
\end{figure*}

\begin{figure*}
 \vspace{0cm}
 \hbox{\hspace{0.5cm}
  \epsfxsize=8.0cm \epsfbox{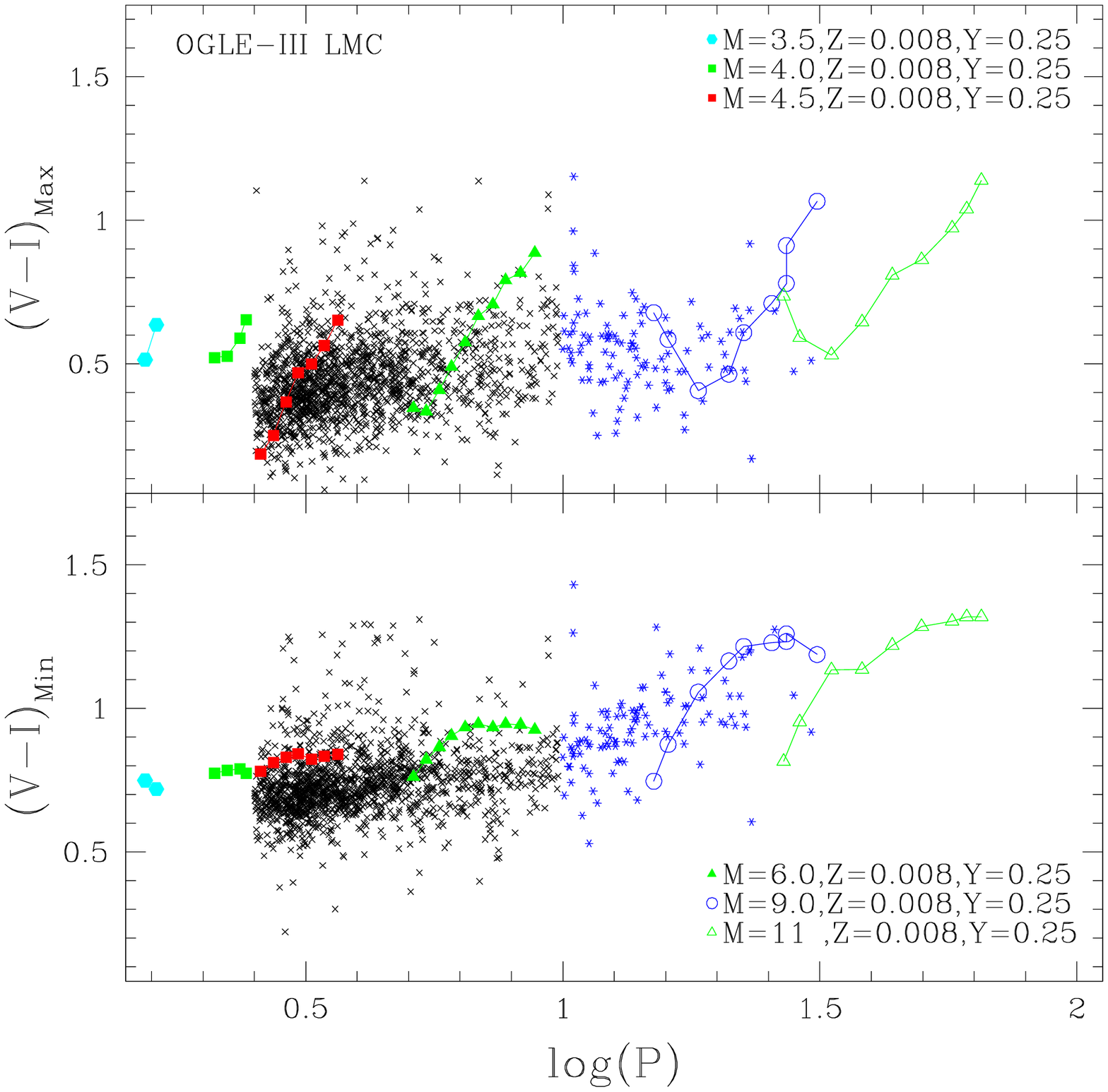}
  \epsfxsize=8.0cm \epsfbox{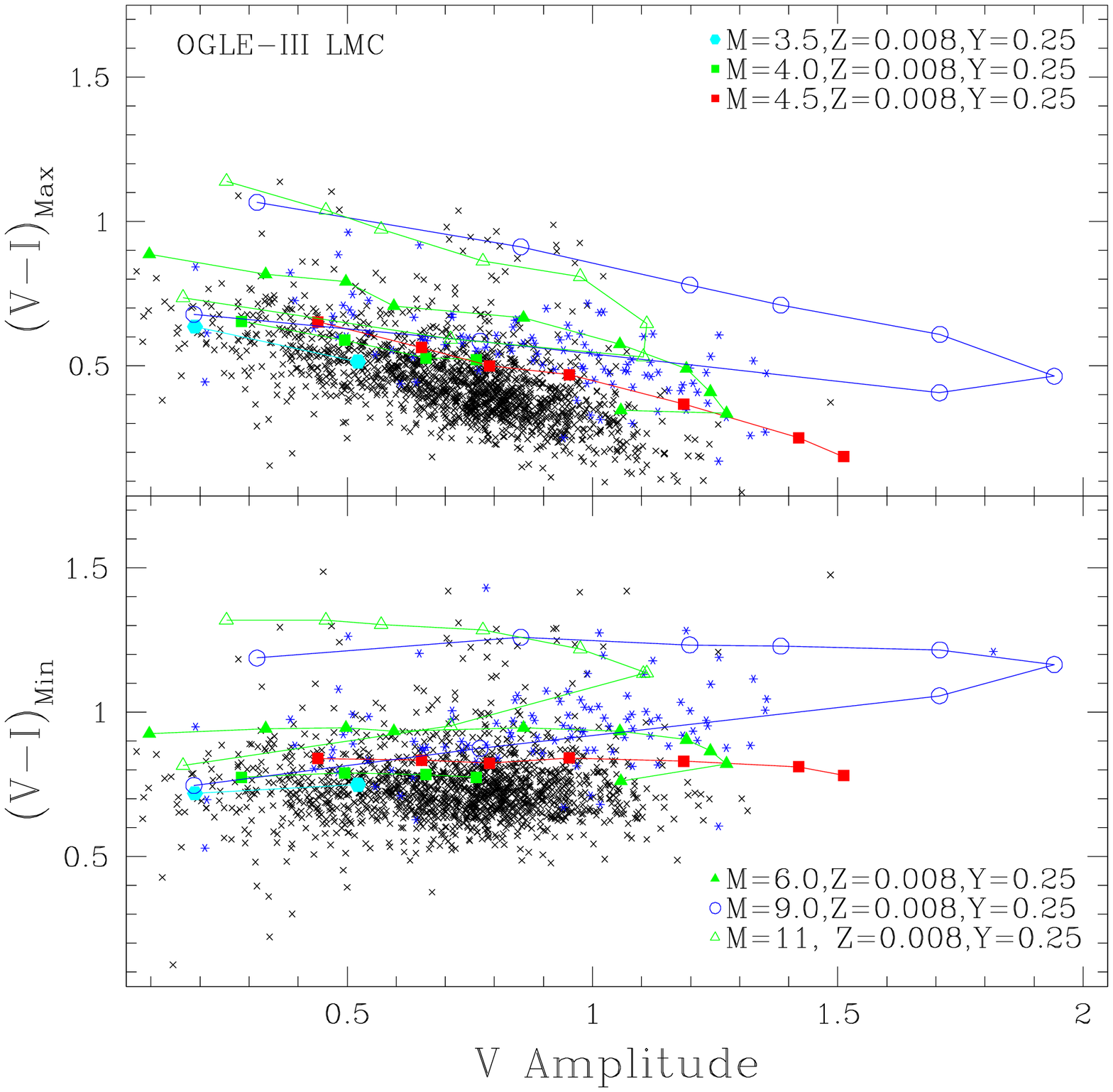}
 }
 \vspace{0cm}
  \caption{Multiphase PC/AC relations for reddening corrected OGLE III LMC Cepheid data against theoretical models with $Z=0.008, Y=0.25.$}
  \label{UNIQUE_LABEL}
\end{figure*}

\begin{figure*}
  \vspace{0cm}
  \hbox{\hspace{0.5cm}
    \epsfxsize=8.0cm \epsfbox{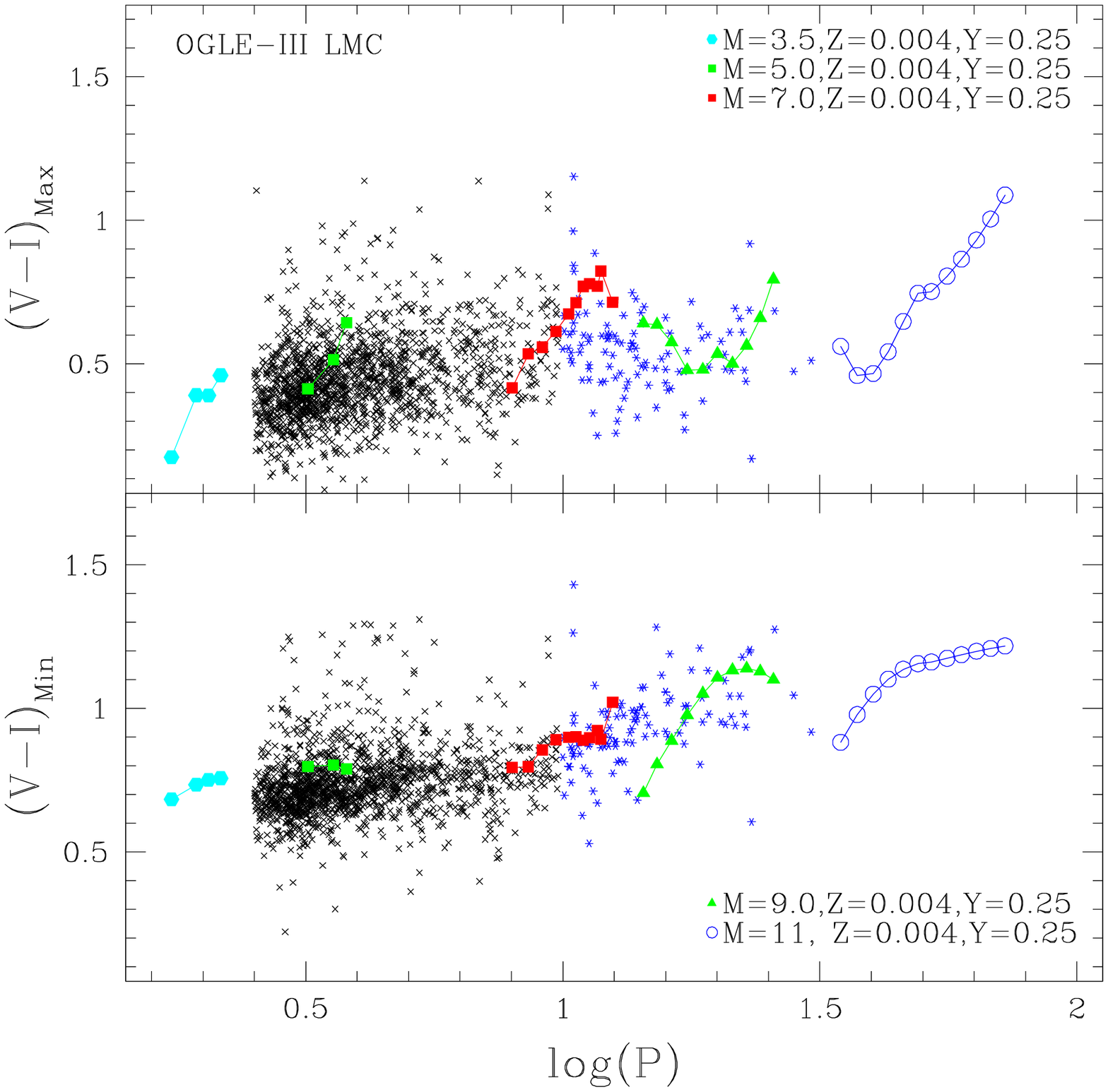}
    \epsfxsize=8.0cm \epsfbox{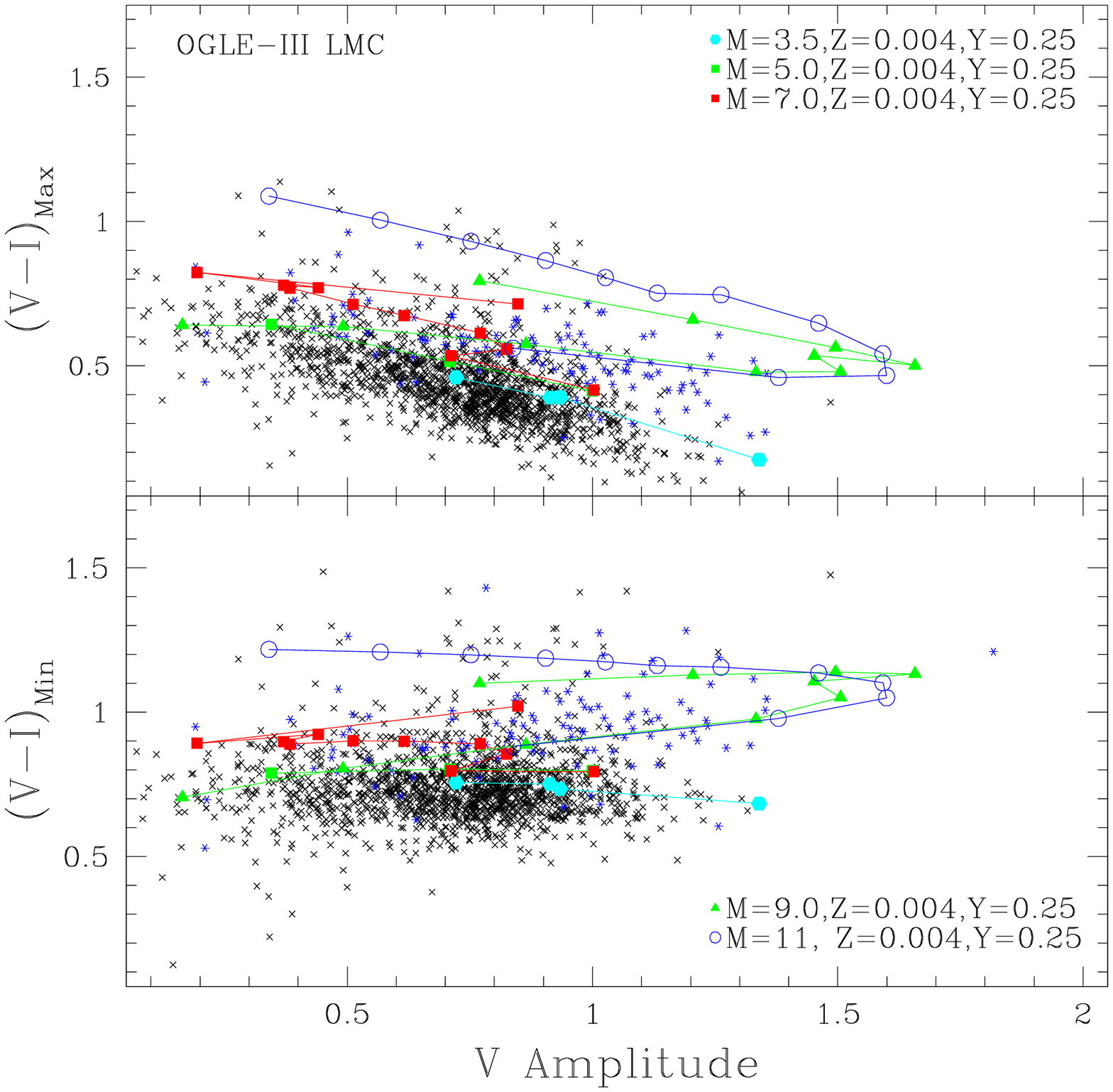}
  }
  \vspace{0cm}
  \caption{Multiphase PC/AC relations for reddening corrected OGLE III LMC Cepheid data against theoretical models with $Z=0.004, Y=0.25.$}
  \label{UNIQUE_LABEL}
\end{figure*}

\begin{figure*}
  \vspace{0cm}
  \hbox{\hspace{0.5cm}
    \epsfxsize=5.5cm \epsfbox{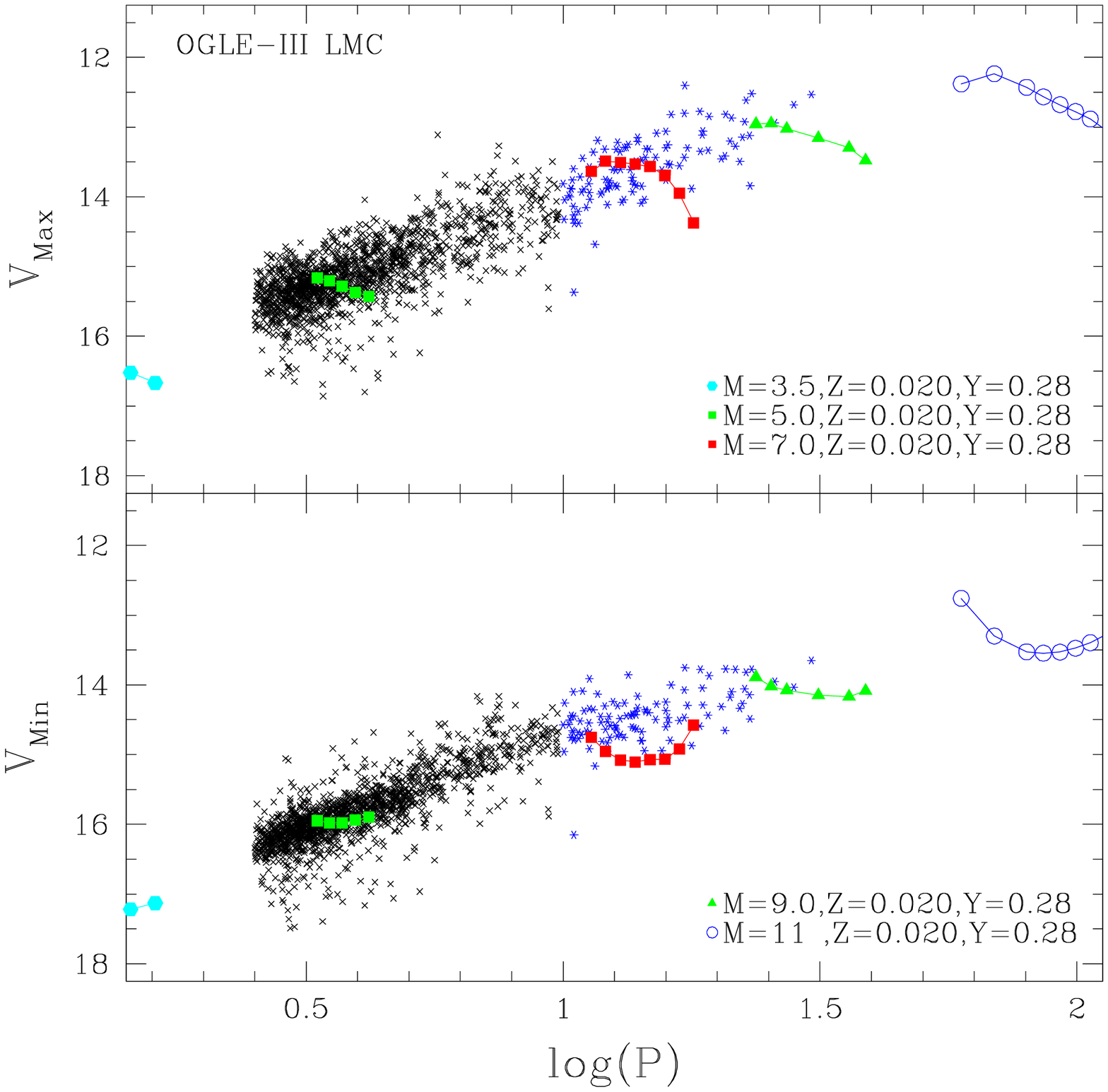}
    \epsfxsize=5.5cm \epsfbox{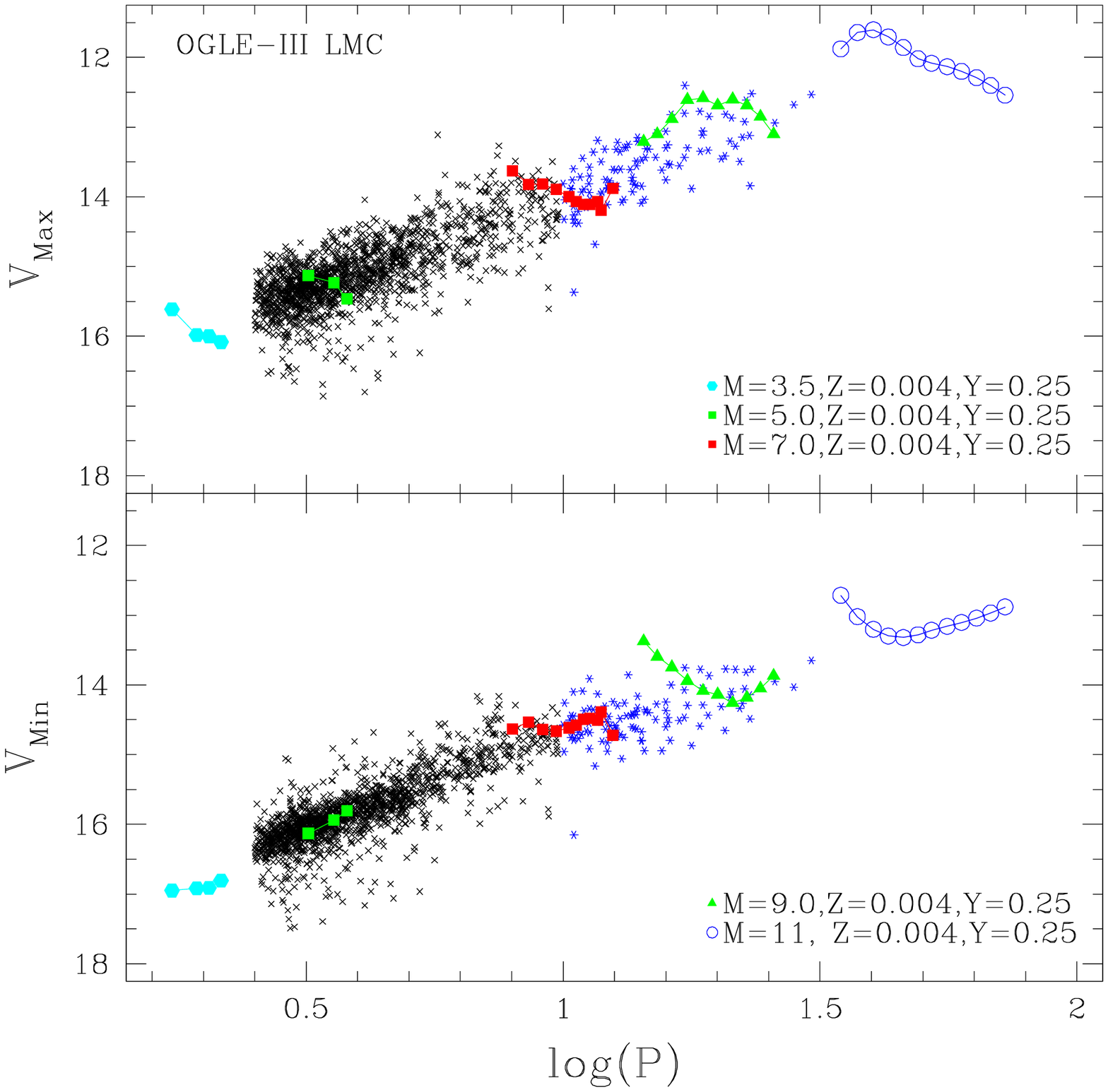}
    \epsfxsize=5.5cm \epsfbox{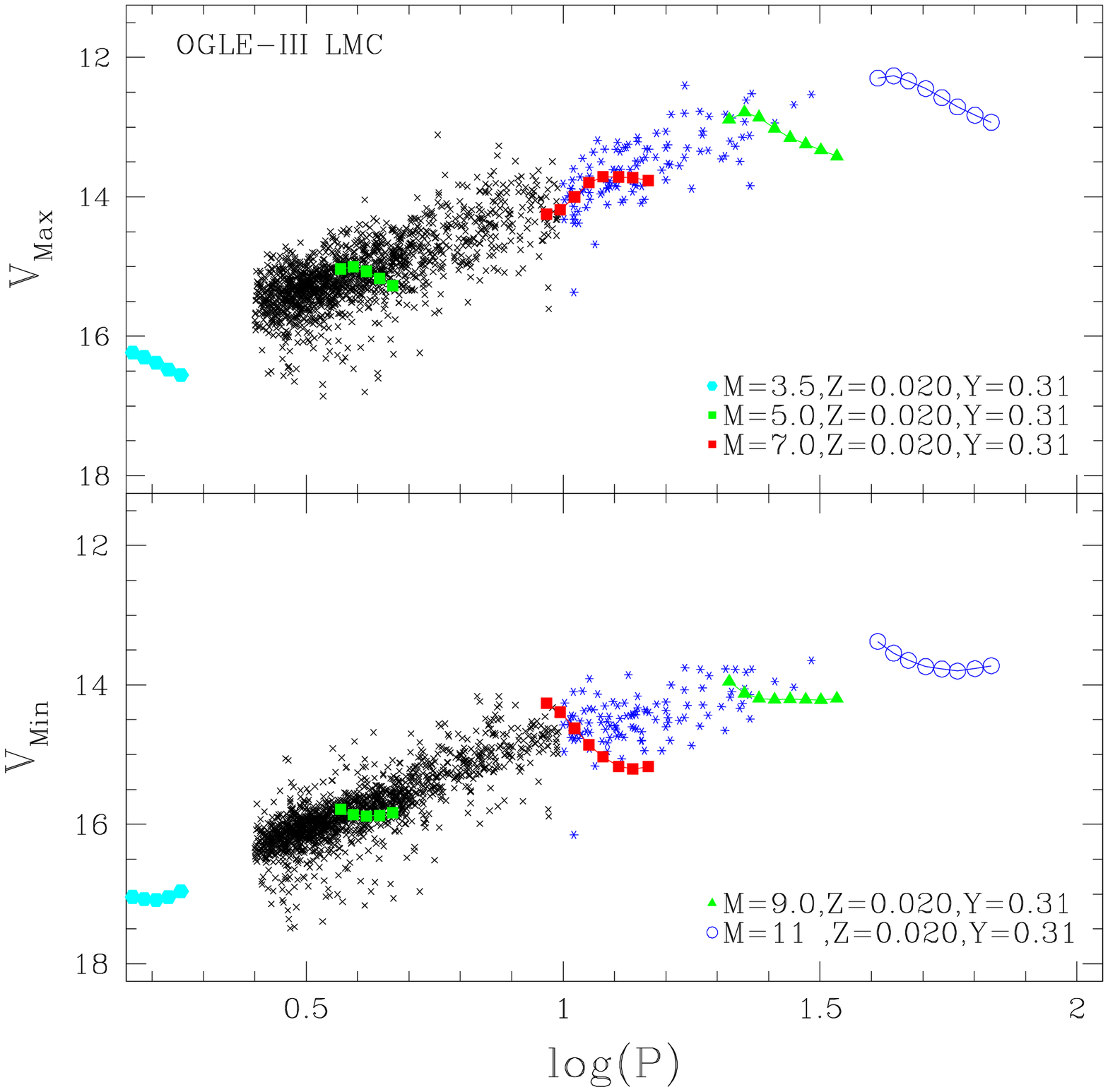}
  }
  \vspace{0cm}
  \caption{ Multiphase PL relations for reddening corrected OGLE III LMC Cepheid data against theoretical models with $(Z=0.02,Y=0.28), (Z=0.004,Y=0.25),
(Z=0.02,Y=0.31).$}
  \label{UNIQUE_LABEL}
\end{figure*}

\begin{figure*}
   \vspace{0cm}
\centering
    \epsfxsize=8.0cm{\epsfbox{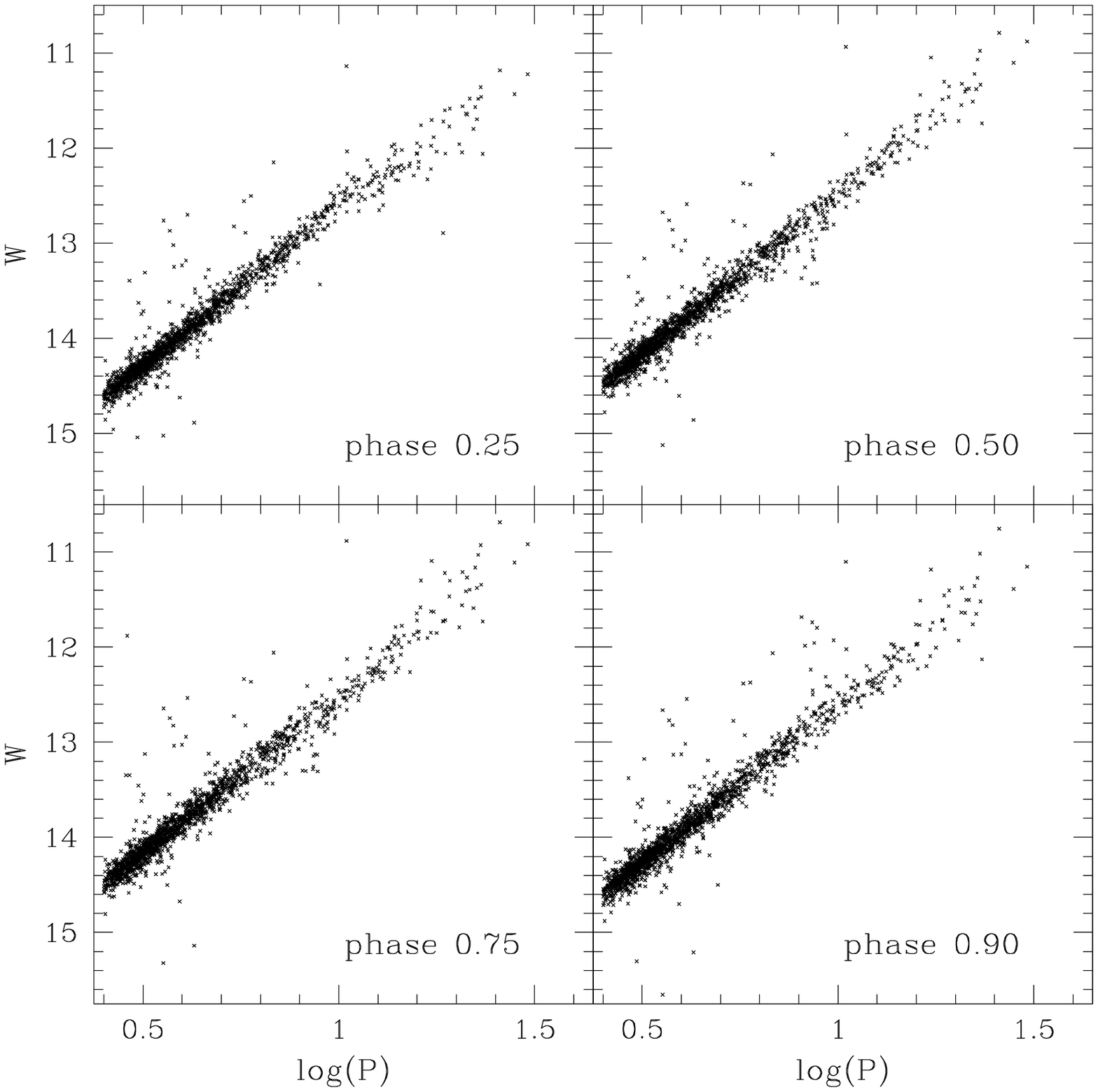}}
   \caption{Multiphase Wesenheit function for OGLE III LMC Cepheid data showing cleanr nonlinearities at certain phases.}
\end{figure*}

\label{lastpage}


\begin{thebibliography}{99}




\bibitem[\protect\citeauthoryear{Bono et al}{2008}]{b08} Bono, G., Caputo, F., Fiorentino, G., Marconi, M., Musella, I., 2008, ApJ, 684, 102

\bibitem[\protect\citeauthoryear{Bono, Marconi \& Stellingwerf}{1999}]{bms99} Bono, G., Marconi, M., Stellingwerf, R.F. 1999, ApJS, 122, 167

\bibitem[\protect\citeauthoryear{Bono et al}{2000a}]{b00} Bono, G., Caputo, F., Cassisi, S., Marconi, M., Piersanti, L., Tornamb\`e A. 2000, ApJ, 529, 293


\bibitem[\protect\citeauthoryear{Bono et al}{2000b}]{bcm00} Bono, G., Castellani, V., Marconi, M. 2000, ApJ, 529, 293

\bibitem[\protect\citeauthoryear{Bono \& Stellingwerf}{1994}]{bs94} Bono, G., Stellingwerf, R.F. 1994, ApJS, 93, 233

\bibitem[\protect\citeauthoryear{Caputo, Marconi, \& Musella}{2000}]{cmm00} Caputo, F., Marconi, M., Musella, I. 2000, A\&A, 354, 610

\bibitem[\protect\citeauthoryear{Castelli, Gratton, \& Kurucz}{1997a}]{cgk97a} Castelli, F., Gratton, R. G., \& Kurucz, R. L. 1997a, A\&A, 318, 841

\bibitem[\protect\citeauthoryear{Castelli, Gratton, \& Kurucz}{1997b}]{cgk97b} Castelli, F., Gratton, R. G., \& Kurucz, R. L. 1997b, A\&A, 324, 432

\bibitem[\protect\citeauthoryear{Freedman et al.}{2001}]{fre01} Freedman, W. L. et al. 2001, ApJ, 553, 47

\bibitem[\protect\citeauthoryear{Kanbur \& Ngeow}{2004}]{kn04} Kanbur, S. M., Ngeow, C., 2004, MNRAS, 350, 962

\bibitem[\protect\citeauthoryear{Kanbur et al.}{2004}]{knb04} Kanbur, S. M., Ngeow, C., Buchler, J. R., 2004, MNRAS, 354, 212

\bibitem[\protect\citeauthoryear{Kanbur et al.}{2006}]{kn06} Kanbur, S. M., Ngeow, C., 2006, MNRAS, 369, 705

\bibitem[\protect\citeauthoryear{Keller et al.}{2006}]{kw06} Keller, S. C., Wood, P. R., 2006, ApJ, 642, 834

\bibitem[\protect\citeauthoryear{Koen et al.}{2007}]{k07} Koen, C., Kanbur, S. M., Choong, C., 2007, MNRAS, 380, 1440

\bibitem[\protect\citeauthoryear{Marconi, Musella, \& Fiorentino}{2005}]{mmf05} Marconi, M., Musella, I., Fiorentino, G. 2005, ApJ, 632, 590

\bibitem[\protect\citeauthoryear{Marconi}{2009}]{m09} Marconi, M., 2009, MemSAI, 80, 141

\bibitem[\protect\citeauthoryear{Natale, Marconi, \& Bono}{2008}]{n08} Natale, G., Marconi, M., Bono, G., 2008, ApJL,  674, 93

\bibitem[\protect\citeauthoryear{Ngeow \& Kanbur}{2004}]{nk04}  Ngeow, C.C., Kanbur, S. M. 2004, MNRAS, 349, 1130

\bibitem[\protect\citeauthoryear{Ngeow et al.}{2005}]{n05}  Ngeow, C.C., Kanbur, S. M., Nikolaev, S., Buonaccorsi, J., Cook, K. H., Welch, D. L. 2005, MNRAS, 363, 831

\bibitem[\protect\citeauthoryear{Ngeow \& Kanbur}{2005}]{nk05}  Ngeow, C.C., Kanbur, S. M. 2005, MNRAS, 360, 1033

\bibitem[\protect\citeauthoryear{Ngeow \& Kanbur}{2006a}]{nk06}  Ngeow, C.C., Kanbur, S. M. 2006a, ApJL, 642, 29

\bibitem[\protect\citeauthoryear{Ngeow \& Kanbur}{2006b}]{n06}  Ngeow, C.C., Kanbur, S. M. 2006b, MNRAS, 369, 723

\bibitem[\protect\citeauthoryear{Ngeow et al.}{2009}]{n09}  Ngeow, C.C., Kanbur, S. M., Neilson, H.R., Nanthakumar, A., Buonaccorsi, J. 2009, ApJ, 693, 691

\bibitem[\protect\citeauthoryear{Romaniello et al.}{2005}]{r05} Romaniello, M., Primas, F., Mottini, M., Groenewegen, M.A.T., Bono, G., François, P. 2005, A\&A, 429, 37

\bibitem[\protect\citeauthoryear{Romaniello et al.}{2008}]{r08} Romaniello, M., Primas, F., Mottini, M., Pedicelli, S., Lemasle, B., Bono, G., François, P., Groenewegen, M.A.T., Laney, C.D. 2008, A\&A, 488, 731

\bibitem[\protect\citeauthoryear{Saha et al.}{2001}]{sa01} Saha, A. et al. 2001, ApJ, 562, 314

\bibitem[\protect\citeauthoryear{Stellingwerf}{1982}]{s82} Stellingwerf, R.F. 1982, ApJ, 262, 339

\bibitem[\protect\citeauthoryear{Simon et al.}{1993}]{skm93} Simon, N. R., Kanbur, S. M., Mihalas, D., 1993, ApJ, 414, 310

\bibitem[\protect\citeauthoryear{Udalski et al.}{1999}]{ud99} Udalski, A., et al, 1999, AcA, 49, 201 



 
\end{thebibliography}
\end{document}